\documentstyle[12pt,epsfig]{article}
\def \S {{\cal S}}

\def \eqref{\ref}
\def \ci {\cite}

\def\be{\begin{equation}}
\def\ee{\end{equation}}

\def \half{{1\over 2}}
\def \ov {\over}

\def \a {\alpha}

\def \ha {{1 \ov 2}}

\def \inv {^{-1}}

\def \inv {^{-1}}

\def \s {\sigma}

\def \ha {\half}
\def \ov {\over}

\def \a {\alpha}

\def\vp {\varphi}

\def \g {\gamma}

\def \r {r} 
 
\def   \td {\tilde }

\def\q {b}
\def\b{\td b}
\def \be {\begin{equation}}
\def \ee {\end{equation}}
\def \la{\label}
\newcommand{\rf}[1]{(\ref{#1})}

\begin{document}

\begin{titlepage}

\vspace{3cm}

\begin{center}
{\LARGE 
Magnetic backgrounds and tachyonic instabilities
\\[.1cm]
    in closed string theory }\\[.2cm]
\vspace{1.1cm}
{\large  A.A. Tseytlin\footnote{\ Also at Blackett Laboratory,
Imperial College, London and   Lebedev Physics Institute, Moscow.} }

\vspace{18pt}
{\it
 Department of Physics,
The Ohio State University  \\
Columbus, OH 43210-1106, USA\\
}

\end{center}

\vspace{1cm}

\begin{abstract}
We consider closed superstrings in  Melvin-type magnetic backgrounds.
A 2-parameter class of such  NS-NS backgrounds
are exactly solvable  as weakly coupled string models 
with  spectrum  containing tachyonic modes. 
Magnetic field allows one  to  interpolate
between free superstring theories
with periodic and antiperiodic boundary conditions
for the fermions around some compact direction, and,
in particular, between type 0 and type II string theories.
Using  ``9--11" flip, this interpolation can be  extended to M-theory
and may be used to study the issue of tachyon condensation 
in type 0 string theory. We review related  duality proposals, 
and, in particular, 
  suggest
a description of type 0 theory in terms of  M-theory
 in a curved   magnetic flux background
in which  the type 0 tachyon   appears to  correspond
 to a state in  $d=11$ supergravity fluctuation spectrum. 
\end{abstract}

\vspace{2cm}

\begin{center}
{\it Contribution to the Proceedings 
     of the     10-th Tohwa International Symposium 
     on String Theory, Fukuoka,  Japan,  July 3-7, 2001.
}
\end{center}

\end{titlepage}
\setcounter{page}{1}
\renewcommand{\thefootnote}{\arabic{footnote}}
\setcounter{footnote}{0}

\section{Introduction}
Magnetic backgrounds play an important role in  field   theory 
and open string theory providing  simple solvable models 
with  nontrivial  physics content. 
Similar (approximately) constant magnetic backgrounds in 
 gravitational theories  
like closed string theory are necessarily curved  with an example 
provided  
   by  the Melvin-type  flux tube solutions (see, e.g., \ci{gib}).
The role of the magnetic field(s) is played by the
vector(s) originating from the metric and/or 
antisymmetric 2-tensor upon  Kaluza-Klein 
reduction on a circle. 

One important special case is the Kaluza-Klein Melvin 
background which has  curved magnetic universe interpretation 
in 9 dimensions, but, remarkably,  is represented by a flat (``twisted'')
10-dimensional space \ci{gaun}. The non-trivial 3-dimensional 
part of this space is a twisted product of a K-K circle and a 2-plane:
going around the circle  must be  accompanied by a rotation by angle $2\pi b$
in the plane. This is a continuous version of the Rohm model \ci{rohm}
where the 2-plane was replaced by a 2-torus so that  the magnetic (``twist'') 
parameter was allowed to take only  discreet values. 

 The  flatness of the 10-d space  allows one 
to solve the corresponding (super)string theory  exactly \ci{magnetic,closed}
determining the ``Landau-type'' spectrum of states
which (for large enough magnetic field) contains tachyons 
in the winding sector.
Other NS-NS Melvin-type models (which are no longer flat in 10 dimensions)
are also solvable since they may be formally related to the K-K Melvin 
model by a generalized 
$O(d,d)$ T-duality transformation \ci{more}.

One can  also construct similar 
magnetic backgrounds with  R-R  magnetic field by applying U-duality transformation, or equivalently, by lifting, e.g.,  the K-K Melvin
space to 11 dimensions and then 
reducing down to 10 along a different dimension
\ci{green,our}. Such R-R string models seem, however, 
hard to solve exactly.

One characteristic feature  of these magnetic closed string models
is that the magnetic field parameter introduces an effective phase for 
space-time fermions; in particular, switching on  magnetic field 
allows one to interpolate between periodic and antiperiodic boundary conditions 
for the fermions along a spatial circle \ci{magnetic,green}.
Antiperiodic  fermions appear in the context of 
superstrings at finite temperature \ci{atick}
and  in closely related 
 type 0 string theory viewed as $(-1)^{F_s}$ orbifold of 
type II superstring theory \ci{typ,BD}. 

Using the orbifold interpretation of type 0 string theory 
and ``9-11'' flip  it was suggested in 
\ci{BG} to interpret type 0 string theory as a similar
non-supersymmetric  
orbifold compactification of 11-d M-theory.  It was  
conjectured    \ci{BG}
 that the tachyon of type 0A string in flat space
  gets  $m^2>0  $ at strong coupling 
where type 0A string
    becomes dual to
  11-d M-theory on  large circle with antiperiodic fermions.

In an interesting paper 
  \ci{CG}, this   proposal was combined 
  with the observation
 that K-K   Melvin 
  magnetic background 
  in type II string theory  (and 
its direct lift to 11 dimensions) 
   may be used to
  interpolate \ci{magnetic,green} between
  periodic and antiperiodic boundary conditions  for
  space-time fermions. Since reducing the flat 
  twisted 11-d space  along a  different --
  ``mixed circle'' -- direction gives 
\ci{gaun,green} 
the R-R  7-brane background, it was 
suggested that type 0 string in this 10-d   R-R  Melvin 
background should be dual to type II theory 
in the same background  but with shifted 
magnetic field.  This implies
  that  type 0 string  tachyon
  should disappear at strong  magnetic R-R field
   \ci{CG,GS}.

In our  recent work 
 \ci{our}, which will be reviewed below, 
we extended the discussion  in \ci{green,CG}
to the case of more general class of  Melvin-type backgrounds 
with two independent magnetic parameters $\q$ and $\b$
(parametrizing the vector and axial vector 9-d fields
originating from $G_{i9}$ and $B_{i9}$ components).
This class   includes the K-K Melvin ($\q\not=0, \b=0)$
 and the dilatonic Melvin ($\q = \b)$ 
 solutions
 as special cases and  is covariant under T-duality.
We  lifted these type IIA magnetic  solutions
to 11 dimensions,  getting a $\b\not=0$
generalization of the flat \ci{gaun}
$d=11$ background discussed  in \ci{green,CG}.
 Dimensional reduction along different directions (or U-dualities
 directly in $d=10$) led to a number of $d=10$ supergravity
 backgrounds with
 R-R magnetic fluxes, generalizing  to $\b\not=0$
 the R-R magnetic  flux 7-brane \ci{green,CG}
  of type IIA theory.
  
We  proposed  the analog of the type II
 -- type 0 duality
in \ci{CG}  now  based on a curved  $d=11$
background which is a lift of the T-dual  $\b \not=0, b=0$
Melvin background. Remarkably, this 
background has manifest 9--11 symmetry, i.e. 
its two different 10-d reductions 
 are formally
 the same NS-NS Melvin spaces.\footnote{Though one corresponds 
 to a weakly-coupled and 
 another to a strongly coupled string theory as
  the values of the radii $R_9$   and $R_{11}$ are interchanged.} 
While in the original K-K Melvin set-up the  tachyon originates from a stringy winding mode
(i.e. type 0 tachyon may be interpreted as corresponding  to 
   a wound membrane state) 
in this T-dual setting  the  tachyon 
appears to be a momentum  mode,  i.e. 
  the type 0 tachyon   appears to  correspond
 to a state in  $d=11$ supergravity multiplet.

 One  motivation  for the study of these 
  magnetic backgrounds and their instabilities 
 is to get 
  better  understanding of closed string tachyons
and their stabilization, i.e. 
of dynamical  interpolations between stable and unstable
 theories.
In particular, 
one would like to  establish  precise relations
  between  non-supersymmetric
   backgrounds in  type II superstring
   theory  and  unstable backgrounds in  type 0 theory.
   Another is to  use  interpolating magnetic backgrounds
   to connect   D-brane solutions
 in the two theories and related    gauge theories.
  Understanding the fate of closed string tachyon 
  in such non-supersymmetric situations 
is crucial, in particular, 
 for the program of constructing string theory duals  
 of non-supersymmetric  gauge theories
\ci{sas,KT,TZ,IG} (see also \ci{cost,adams,ada}   for 
some recent related discussions).

\section{NS-NS Melvin background in $d=10$ superstring }
Consider the following NS-NS bosonic background 
in type II string theory:
\be \la{melv}
ds^2=-dt^2 + dx_s^2+dx_9^2+dr^2+ {r^2\over 1 +\b ^2 r^2}  [d\varphi
+(\q +\b  ) dx_9] [d\varphi +
(\q -\b  ) dx_9]\ ,
\ee
\be\la{meel}
B_2 ={\b  r^2\over 1+\b  ^2 r^2} d \vp \wedge dx_9 \ , \ \ \ \ \ \ \ \
\ \ \ \
e^{2(\phi -\phi_0)}={1\over 1+\b ^2 r^2}\ . 
\ee
This is an exact solution of the string theory
-- the corresponding 
 sigma model is conformal  to all orders in $\alpha'$ \ci{magnetic,closed}.
 $x_9$ is a periodic coordinate of radius $R=R_9$ and  $\vp $
 is the  angular  coordinate with period
 $2 \pi$.  This model is covariant under T-duality in $x_9$ direction: the 
 $(R, \b ,\q )$ model
is  T-dual ($x_9 \to \td x_9$)
 to $(\td R, \q , \b )$ model with
$
\td R\equiv {\a'\over R}$. 

 The constants $\q$ and $\b$ are the magnetic field
 parameters.
The magnetic Melvin-type flux tube
 interpretation becomes
apparent  upon dimensional reduction in the $x_9$-direction.
The resulting solution of $d=9$ supergravity
\be\la{backg}
 ds^2_9=  -dt^2 + dx_s^2   + dr^2 +   { f^{-1}  \tilde f^{-1}}
r^2 d\vp^2  \ ,\
\ee
\be\la{baag}
 {\cal A}_\vp=  \q       r^2 f\inv  \ , \ \ \ \ \ \ \
{\cal B}_\vp=  -\b     r^2 \td f\inv  \ , \ee
\be\la{deq}
e^{2(\phi-\phi_0)}=\td f^{-1} \ ,\ \   e^{2\s} =
 f  \td f \inv
 \ , \ \ \ \ \ \ \ \ \ \ \
  f  \equiv 1 +  \q ^2  r^2  \  ,\ \ \
\tilde f  \equiv  1 + \b  ^2  r^2 \ , \ee
 describes  a magnetic flux tube universe with
 the magnetic  vector ${\cal A}$ (coming from the metric)
  and  the axial-vector $ {\cal B}$ (coming from the 2-form)
and the non-constant
dilaton $\phi$ and K-K  scalar $\s$.
 
There are two special important cases:
(i) $\b=0, b\not=0$ and (ii) $b=0, \b\not=0$.  
The corresponding string backgrounds
 are T-dual to each other. 
The first (K-K Melvin model) 
is represented by the flat  10-d space
(but curved magnetic 9-d background) 
\be\la{fff}
ds^2=-dt^2 + dx_s^2+dx_9^2+dr^2+
r^2  (d\varphi +\q\  dx_9)^2  \ .
\ee
The second 
has non-trivial metric, 3-form and dilaton:
\be  \la{beb}
 ds^2_{10}=  -dt^2 + dx_s^2   + d\r^2 + \td f^{-1} (dx_9^2 +  {\r^2 } d\vp^2)  \ .
\ee   \be \la{kok} 
H_3=dB_2 =2 \b   \td f^{-2} r dr \wedge  d \vp \wedge dx_9 \ , \ \ \ \ \ \ \ \
\ \ \ \
e^{2(\phi -\phi_0)}=\td f^{-1} \ . 
\ee
They define equivalent conformal field theories, 
with the two different geometries  being ``seen'', as in other known T-dual
situations,     by 
point-like momentum and  winding string modes respectively.
\footnote{While the curved 2-parameter background \rf{melv} may look much more
complicated than the flat one \rf{fff}, it is the natural 
T-duality (O(2,2) duality) ``completion'' of \rf{fff}.
A possible analogy is to think of \rf{fff} as a counterpart of a
plane wave  background; then  \rf{beb} corresponds to the
fundamental string background, and \rf{melv} -- to a superposition 
of a fundamental string and a wave.}

For generic values of $\q  $ and $\b  $ all of the type II
supersymmetries are broken; 
 supersymmetries are restored at the
 special values $\q R=2n_1$ and $\b  \td R=2n_2$ ($n_i=0,\pm 1, ...$),
where the conformal model describes the
standard flat-space  type II  superstring theory.

As follows from \rf{fff}, 
a shift of $x_9$ by period of the circle
$2 \pi R$  implies rotation in the plane by angle $ 2 \pi b R $.
Thus in  the special case of $b R=2k +1 $ ($k=0, \pm 1, ...$)
the metric becomes topologically trivial. However,
the superstring theory
in this 
 background is still non-trivial (not equivalent to
that of standard flat space)  since space-time fermions
change sign under $2\pi$ rotation in the plane. That means that the $b R=1$
case (all models with $b R=2k +1$ are equivalent)
  describes superstrings with {\it anti}periodic
boundary condition in $x_9$, just as in the
 twisted 3-torus model        of \ci{rohm}
or in the finite temperature case \ci{atick}. 
 Type IIA string  in the space \rf{fff} with $\q R=1$ is
equivalent, for $R\to 0$,
to type 0A string
 in  $R^{1,8} \times S^1_{R\to 0}$ or type 0B string in
$R^{1,8} \times S^1_{\tilde R\to \infty}$.
 The tachyon of type 0 theory
  originates from  a particular   winding mode
 present in the $\q R=1$  type II model
 spectrum.

The string model corresponding to the above 
2-parameter background can be solved in terms of free fields 
and its spectrum is found to be 
$$
 \ha  \a'  M^2 \equiv
 \ha  \a' ( E^2 -  p_s^2  ) =
 \hat N_R+  \hat N_L +\   \ha{ \td R R\inv }
(m- \q R\hat J)^2
$$ \be\la{hail} 
+ \ \ha { R  \td R\inv }  (w  -  \b \td R  \hat J)^2
-   \g  (\hat J_R-\hat J_L)=0 \ ,
\ee
where 
$\hat N_R-  \hat N_L = mw , $ and 
$
 \g\equiv  \q Rw + \b \td R m
-  \q  R \b \tilde R \hat J \ . $
Here $p_s$  are continuous  momenta in the 6 free directions,
 the integers $m$ and $w$ represent  quantized
 momentum and winding numbers in the compact $x_9$ direction, 
$\hat N_R$ and $\hat N_L$ are the   number of states operators, which
have the standard free string theory
form in terms of normal-ordered bilinears of
bosonic and fermionic oscillators, and 
$\hat J\equiv \hat J_L + \hat J_R $ and
$\hat J_{L,R}$ are the angular momentum operators in the 2-plane
with the
 eigenvalues
$\hat J_{L,R}
 = \pm (l_{L,R} + \ha) + S_{L,R}   ,  \  \
 \hat J= l_L-l_R + S_L + S_R .
$
The orbital momenta $l_{L,R}=0,1,2,...  $ (which replace the continuous
linear momenta $p_1,p_2$ in the $(r,\vp)$ \ 2-plane
for non-zero values of $\g$)
 are  the analogues of the  orbital quantum number $l$ 
 and the radial quantum number $k$  in the Landau problem,
 and $S_{R,L}$ are the  spin components.

The spectrum  is thus similar to the one for charge particles
in constant magnetic field orthogonal to the plane of motion 
(with  $Q_{L,R}=mR^{-1}\pm w \td R\inv  $
playing the role of charges of string states in 9-dimensional
description). In contrast to the open string case, here not only the masses and spins but also charges 
of string states can take arbitrarily large values. 
The terms in $M^2$ that are  linear in angular momenta   reflect gyromagnetic 
interaction, while the terms quadratic in $J$'s  
represent the effect of gravitational back reaction of the 
magnetic field.

Since  $\hat J$ can take
both
integer
(in NS-NS, R-R sectors) {  and}
{ half}-integer  (in NS-R, R-NS sectors) values,
the symmetry of the
bosonic
part of the spectrum  $\q\to \q+ k_1 R\inv, \  \b\to \b+ k_2
 \tilde R\inv $  is { not}
 a   symmetry of  its
fermionic part, i.e. the full  superstring spectrum is
invariant under the shifts 
      with  even integer  $k_i=2n_i$ only.
In the case  of $\q R =2n_1, \ \b \tilde R = 2 n_2 $ (i.e.
$\g=$even integer, $\hat \gamma=0$) the
 spectrum is thus equivalent to that of  the standard  free superstring
compactified on a circle.
In the two cases
$
(a)\ \
 \q R={\rm odd}, \  \ \b  \td R= {\rm even} $ 
 { or }\  \  $
(b)\ \
  \q R={\rm  even},\  \  \b  \td R={\rm  odd } \ 
$
the spectrum
 is
 the same as that
of  the  free superstring compactified
on a circle with {\it anti}periodic boundary conditions
for space-time fermions.

 The T-duality symmetry in the compact Kaluza-Klein direction $x_9$
is manifest in the spectrum:  $M^2$  is
invariant under  $R\leftrightarrow \td R\equiv \a'R\inv ,\
 \q  \leftrightarrow \b,
\ \ m\leftrightarrow w$.
Thus the two T-dual models -- $b=0$ one  and $\b=0$ one --
have   equivalent spectra.

The only states which may  become  {\it tachyonic }
  are bosonic states that  lie on
the first Regge trajectory with maximal value of $S_R$, minimal value of
$S_L$,  and   zero orbital momentum,
i.e.  $\hat J_R=S_R-\ha=\hat N_R+ \ha ,\   $  $\hat J_L=S_L+\ha=-\hat N_L -\ha
$, so that $\hat J_R-\hat J_L= \hat N_R +\hat N_L +1$.
For general $b$ and $\tilde b$,  there are
instabilities (associated with states with  high spin and charge) for
arbitrarily small values
of the magnetic field parameters.
The only exception is the   $\b =0$  model  and  equivalent T-dual
 $\q =0$ model. For   $\b =0$ model  the type II superstring
   has no tachyons if  the value of $\q$ is smaller than
  some {\it finite} critical value $\q _{\rm cr}$.
The spectrum of the $\b =0$ model ($\gamma = \q  R  w $)
has the first (lowest mass)
potentially tachyonic state  that appears
  as $\q $ is increased from zero  for 
$\hat J=0, \ \hat J_R-\hat J_L=1$:  it has 
$
\hat N_R=\hat N_L=0,  \ 
 m=0 ,\  w=1 ,  \ l_R=l_L=0 ,  \ S_R=-S_L=1\ ,
$ and thus 
$$\a' M^2= {R^2\over\a' } - 2\q R \ \equiv  \ 2R ( \q _{\rm cr}  -\q)   \ ,
\ \ \ \ \ \ \   \ \q _{\rm cr}= {R\over 2\a'} \ ,   $$
i.e. we  get tachyon 
in the winding sector 
when    $\q >\q _{\rm cr}$.
In the   T-dual model with $\q =0, \
\b\not=0  $, which has the equivalent
 spectrum, the first tachyon appears in momentum sector 
 for   $\b  >\b _{\rm cr }= {{ \td R} \ov 2 \a'} = {1\over 2R }$.

\section{Related $d=11$ backgrounds and U-dual $d=10$ R-R 
flux branes}
Lifting the type IIA background \rf{melv} to $d=11$ 
one gets the following ``6-brane" (or ``8-brane'' 
with ``defects'' along   $x_9$, $x_11$) 
solution of $d=11$ supergravity
\ci{our} 
\be \la{surr}
ds^2_{11}= \tilde f^{1/3} ( -dt^2+dx_s^2)
 + \tilde f^{-2/3} \big[ dx_{11}^2 +  \tilde f  dr^2  + r^2  f\inv  d\varphi^2
+  {f }\big( d x_9  +{\q r^2 f\inv } d\varphi\big)^2 \big]
\ , \ee
$$
C_3 = \b  r^2 \tilde f^{-1} d x_{9} \wedge d x_{11}
 \wedge d \vp   \ , \ \ \ \ \  f  = 1 +  \q ^2  r^2  \  ,\ \ \
\tilde f  \equiv  1 + \b  ^2  r^2 \ .
$$
This background is regular, with curvature being constant near the
core $r=0$  and going to zero at large $r$. 

The two special cases are  $\b=0$ when the $d=11$ space 
becomes flat (this is the  direct $d=11$ lift of \rf{fff}) 
\be \la{meee}
ds^2_{11}=-dt^2+dx_s^2 + dx_9^2
+dr^2+ r^2 (d\vp +  \q \ dx_9 )^2+ dx^2_{11} \  , \ \ \ \ \ \ 
C_3=0 \ , \ee
and $\q=0$
when the metric and $C_3$ are still non-trivial 
\be \la{surre}
ds^2_{11}= \tilde f^{1/3} \big( -dt^2+dx_s^2  \big)
 + \tilde f^{-2/3} \big( \tilde f dr^2  + r^2    d\varphi^2
 + dx^2_9  +   dx^2_{11} \big)
\ . \ee
$$ C_3 = \b  r^2 \tilde f^{-1} d x_{9} \wedge d x_{11}
 \wedge d \vp   \ . $$
Remarkably,  this metric and $|C_3|$ 
turn out to be  {\it symmetric} under the ``9-11" flip.

The     reduction  of  \rf{surr}
along $x_{11}$ gives back \rf{melv}, 
while the  reduction along $x_9$  leads to the 
following ``mixed''(NS-NS/R-R)  type IIA background
related to \rf{melv} by a U-duality 
(T$_{x_9}$ST$_{x_9}$ duality transformation)
\be\la{dosa}
ds^2_{10}= f^{1/2}\big( -dt^2+dx_s^2+ {\tilde f\inv  }dx_9^2 + dr^2 +
{ f\inv  \tilde f\inv } r^2  d\varphi^2\big)\ , \ee 
$$
A  = \q  r^2 f^{-1} d \vp \ ,\ \ \ \ \ \ \ \ \
 B_2 =\b  r^2 \tilde f^{-1}  {d \varphi\wedge d x_9 } \ ,\ \ \ \ \ \
\ \ e^{2(\phi-\phi_0) }= f^{3/2}\tilde f\inv  \  .
$$
Here we renamed $x_{11} \to x_9$.
The two parameters $\q$ and $\b$ now 
control the strengths of the  magnetic
1-form  R-R  field $A$ and of the  NS-NS  2-form field $B_2$.
This solution may be interpreted as a R-R magnetic flux 
7-brane with a ``defect'' along  $x_9$.\footnote{Note that 
while \rf{melv} had magnetic interpretation in  9 dimensions, 
here one got a magnetic R-R vector directly in 10 dimensions.}

In the special case of $\tilde b =0$ we get the 
 R-R flux 7-brane background \ci{green}  which is the reduction 
 of the flat $d=11$ background \rf{meee} along the other -- $x_9$ --
 direction 
 (and thus is U-dual of the flat K-K Melvin space \rf{fff}
 which is the trivial reduction along $x_{11}$)
\be \la{mccc}
ds^2_{10}=f^{1/2}\big( -dt^2+dx_s^2 + dx_9^2 +
dr^2 + r^2 f\inv d\vp ^2 \big)\ ,
\ee
$$
A_\vp =\q r^2   f\inv \  \ , \ \ \ \ \ \ \
\ \ \  e^{{2}(\phi -\phi_0)}= f^{3/2} \ .
$$
Like in the NS-NS Melvin case \rf{baag}, the 
 strength of the  the R-R 1-form is approximately constant 
near the core, decaying away from the core. However, in contrast to 
\rf{meel}  here the dilaton grows away from the core, so the theory
is weakly-coupled only at small $ r \ll b^{-1} $ \ci{green,CG}.

In the other special case $b=0$ the  $9-11$ symmetry 
of \rf{surre} implies that here the two ``U-dual'' 
reductions are formally  {\it equivalent}, 
leading   to the {\it same}   (pure NS-NS)  background
 \rf{beb} (with the dilaton decreasing away from the core).
 

\section{Perturbative and non-perturbative 
 type 0 -- type II relations}  

The above considerations 
 suggest \ci{CG,our}   non-perturbative dualities 
relations between type II and type 0 strings in magnetic 
backgrounds. 
Let us first review the well-known perturbative flat space
orbifold relation between these  two string theories. 
 
Type 0 closed string theory \ci{typ}  which is 
a ``symmetric'' $(-1)^{F_L+F_R}$ modular-invariant 
projection of the NSR string may be interpreted also as 
 $(-1)^{F_s}$ ($F_s$ is space-time fermion number)
orbifold of type II superstring theory. 
The orbifolding projects out all space-time fermions and 
introduces extra twisted sector states (in particular,  the tachyon and 
another copy of massless R-R bosons).
This 
relation between type II and type 0 string
theories may be expressed by saying that 
type 0 string is (a limit of)  type II string compactified 
on a circle with {\it anti}periodic
 boundary conditions for space-time fermions
\ci{atick}. 
More  precisely, the two theories
are the limits of the same  interpolating  ``9-dimensional"
string 
theory \ci{BD} --  $\Sigma_R$  orbifold of type II theory.
 $\Sigma_R$ stands for $  (S^1)_R/[(-1)^{F_{s} }\times {\cal S}]$,
 where
 $\S$ is half-shift along the circle ($X_9 \to X_9 + \pi R$).\footnote{The
 role of ${\cal S}$ is to mix $(-1)^{F_{s} }$  orbifold
 with compactification on the circle,
  allowing for the  interpolation
  \ci{BD}. Invariant states under ${\cal S}$  have  even momentum quantum
  numbers $m$ and integer winding numbers $w$, while
 twisted sector states  have  odd $m$ and half-integer $w$.
  The action  of $\cal S$ is irrelevant for $R\to \infty$, but
  is crucial for reproducing type 0 spectrum for $R\to 0$.}
 Type IIA on $\Sigma_{R\to \infty}$  is
 type IIA theory  in flat $d=10$ 
  and  type IIA on $\Sigma_{R \to 0}$ is
  type 0A theory  on $(S^1)_{R\to 0}$ (or T-dual  type 0B theory
 on  $(S^1)_{ R\to \infty}$). Thus the $\Sigma_R$ orbifold 
  type IIA  theory  (which has massive fermions 
  in its spectrum) 
  continuously 
  interpolates between  supersymmetric type IIA
 and non-supersymmetric 
 type 0B {\it ten}-dimensional theories \ci{BD}.
 
It is possible to replace the orbifolding procedure 
 by the K-K Melvin background  with $bR=1$.
Indeed, 
 using that type II superstring
in the  $d=10$ K-K    Melvin  NS-NS background \rf{fff}
(with $x_9 \equiv x_9 + 2 \pi R$)
is,   for $\q R=1$,  equivalent
  to type II theory on $R^9 \times (S^1)_R$
  with antiperiodic boundary
 conditions for the  space-time fermions
 along the $x_9$ circle  \ci{magnetic},\footnote{This
  model at $\q R=1$ is
 essentially  the same as  a
   limit ($T^2 \to  R^2$)  of  the
 twisted 3-torus $T^2 \times S^1$
  model of \ci{rohm} 
   or a Wick rotation of the
 finite temperature superstring  theory \ci{atick}, 
 and is closely related to Scherk-Schwarz  compactification 
 in  string theory \ci{KR} (see also \ci{KO}.}
 one may   notice  \ci{CG} that
 it is  equivalent  to  the  above orbifold  of
 type II on $\Sigma_{R'}$   with  the
  radius $R'= 2 R$ (extra  2 is related to the  shift
   ${\cal S}$).\footnote{In the Melvin model  with
   $\q R=1$ the shift  $x_9 \to x_9 + 2\pi R$
   is equivalent to $2\pi$ rotation in the 2-plane under which
   space-time fermions  change sign. In the orbifold,
   the corresponding shift  $\cal S$ is  by $ \pi R'$.}
  This  NS-NS Melvin model thus 
 describes, in the limit $R\to 0$, the 
  weakly coupled  type 0 string on  $R^{1,8}\times
 (S^1)_{R\to 0}$ (or type 0B on $R^{1,8}\times
 (S^1)_{R\to \infty}$).
 
 To summarize, the type II orbifold on $\Sigma_{2R}$ 
  is the same as
 type II theory in Melvin background at $\q R=1$ and
 thus the latter model  interpolates between type II
 and type 0 theories in  infinite $d=10$ space.
In general \ci{CG},  type IIA(B) theory
 in the Kaluza-Klein   NS-NS  Melvin background
 with parameters $(\q , R)$
  is equivalent  
 (has the same perturbative spectrum and thus the 
  same torus partition
 function)  to the   orbifold of
 type IIA(B)  on $\Sigma_{R'}$
  in this  NS-NS  Melvin background with
  parameters
$(\q'= \q - R^{-1},\ R'= 2R)$.
Since the K-K Melvin model \rf{fff} has the same  
spectrum as the T-dual model \rf{beb}, 
a similar statement is true also for the 
type II and type 0 strings in this dual curved  background
\ci{our}.

 Returning  now  back to the case of the trivial 
 flat background, the above
 compactification of type IIA theory on $\Sigma_{R}\equiv \Sigma_{R_9}$ 
  corresponds
  to M-theory
 on $(S^1)_{R'_{11}} \times (S^1)_{R_{9}}/[(-1)^{F_{s}
 }\times {\cal S}]$, where
 $R_9$ and $R'_{11}$
 are  taken to be small.
  Making the ``9-11'' flip (i.e. exchanging the roles
  of the two circles,  assuming  that there is indeed
   an interpolating
 coordinate-invariant 11-d  M-theory)
 one  may  then conjecture \ci{BG}
 that one should  get an equivalent
 description of this  theory
 as an ordinary (i.e. with  periodic fermions)  $S^1$ 
compactification  of the $d=10$ theory obtained from M-theory on
$\Sigma_{R'_{11}}$ (in the limit $R'_{11}\to 0$). 
 Then
type 0A  theory may be interpreted as
   M-theory on $
   \Sigma_{R'_{11}}$, or, essentially,
   as M-theory on  $S^1$ with  radius
   ${ 1 \ov 2}  R'_{11}$ with periodic
   boundary conditions for the bosons and 
     antiperiodic for the 
    fermions.\footnote{
 The factor of $1/2$ is
   again due to the half-shift $\S$. It  suggests that 
    the same  factor should be  present 
   in the relation between string coupling constants.
   However,  it is not clear how  to decide this 
   unambiguously since the relation to perturbative 
    type 0 theory based on 9-11 
   flip applies only in the limit of zero radius, i.e. at 
   zero coupling.}


Lifting the flat K-K   Melvin metric \rf{fff}  to 11 dimensions,
replacing $x_9 \leftrightarrow x_{11}$  in \rf{meee}, 
and reducing it down to
10 dimensions  along $x_{11}$  gives a   non-supersymmetric
 R-R  Melvin   7-brane in
type IIA theory \rf{mccc}.
Combining this fact with the above observations, it was 
suggested in \ci{CG}
that type IIA  and type 0A theories  in the  $d=10$
 R-R Melvin flux 7-brane background \rf{mccc}\footnote{Any 
  bosonic solution
 of type II supergravity can be embedded into type 0 theory
 provided the fields of the twisted sector (tachyon and
 second set of R-R bosons) are set equal to zero \ci{KT}.}
   are non-perturbatively
   dual to each other, with parameters  related by
 $ \q_0= \q- R^{-1}_{11}, \ g_{s0} 
= { R'_{11}\ov   2 \sqrt { \a'} }
 ={ R_{11}\ov   \sqrt { \a'} }$.
 For example,  for $\q R_{11}=1$ the $d=11$  Melvin theory  has  
  antiperiodic fermions on the   circle $R_{11}$, while, according 
  to \ci{BG}, type 0 theory  is M-theory with antiperiodic 
  fermions on circle $\ha R'_{11}$.
 That means, in particular, that starting with type 0 theory
 and increasing  the  value of the R-R magnetic
 parameter $\q_0$, the tachyon should disappear at strong
 enough magnetic field \ci{CG}  when the theory
  becomes strongly coupled --  its
 weakly-coupled description  is in terms of
  stable weakly coupled type II theory
 in R-R Melvin background with small $\q $.\footnote{As explained in
 detail in \ci{our}, one indication that, as in the 
 weakly-coupled $AdS_5 \times S^5$ example in \ci{KT}, 
 the  R-R flux may 
 shift the value of the tachyon mass 
  follows from 
 the presence of a non-minimal $T^2  F^2_{mn} $ coupling 
 \ci{KT} of the tachyon to the R-R background.}

Replacing the starting point of the above  discussion --
K-K Melvin space \rf{fff} (and thus also 
its $d=11$ 
counterpart \rf{meee}) 
by the perturbatively equivalent T-dual background 
\rf{beb} (with the associated $d=11$ background \rf{surre})
we are  able to give an alternative formulation \ci{our}
of the type 
II -- type 0 duality conjecture.
Indeed,  there are
two  magnetic type II  models  which are
equivalent to  type II superstring  theory   in flat space
with fermions obeying {\it anti}periodic
boundary
conditions in the
$x_9$ direction: 
(a) the model with $\b=0,\  \q R=1$ ;\ \
 (b) the model with $\b
\tilde R=1, \  \q=0$.
They are equivalent (T-dual)  as  weakly-coupled  type II
superstring theories compactified on a circle.
 T-duality equivalence is believed to hold also 
 for finite coupling, so one
expects  that M-theory  compactified  along
   $x_9,x_{11}$  in the
  background  \rf{meee}  with $ \q R=1$ and in  \rf{surre}
  with  $\b
\tilde R=1$    should be equivalent not only in the weak-coupling
  limit $R_{11} \to 0$ but also for finite $R_{11}$.
Thus it is natural to conjecture that M-theory in the  background
 \rf{surre}  with $\b \tilde R_{9} =1$  compactified on
 the $(x_9,x_{11})$ torus  with $(+,+)$ (periodic)   boundary conditions
 for the fermions is equivalent to
 M-theory in flat space compactified
 on the 2-torus  with the  $(-,+)$ fermionic  boundary
 conditions. One may thus propose  the following 
 description of type 0A string theory:  
 as M-theory in  the
background
\rf{surre}   with  periodic $x_9$ and
$\b \tilde R_{9} =1$.
 Because of the 9-11 symmetry
 of \rf{surre}
  one may  consider   also
the equivalent
model with $\b \td R_{11}=1$.

While the reduction of  \rf{meee} to 10 dimensions 
gives  string theory in R-R F7 background,  the reduction  of 
\rf{surre}  gives apparently much simpler 
string theory in the  NS-NS background 
\rf{beb} 
the spectrum of which (in the  {\it weakly-coupled} regime)
   we know. 
However,  the 9-11 flip  necessary to  relate the 
 type II and type 0 theories implies that, e.g., a weakly coupled
($R_{11} \ll R_{9}$) type II theory is mapped into  a strongly
coupled  ($R_{9} \ll R_{11}$)  type 0 theory   or  vice versa.
That  precludes one from drawing immediate conclusions 
about the  presence or absence of tachyons  in 
type 0 theory in the NS-NS Melvin background directly 
from the known weakly coupled string spectrum.\footnote{In
 particular, the weakly coupled type 0 theory in  the  
 NS-NS Melvin background is certainly unstable for any 
value of magnetic field parameter (because of its
own  tachyon  present already 
 in flat space)
but it is expected to become stable at strong coupling 
and critical magnetic field  since   it should then become 
 equivalent  to a weakly coupled type II string in 
NS-NS background with  a small magnetic field parameter.}

According to the proposal of  \ci{CG}, type 
IIA and type 0A theories in
 the $\b=0$  R-R Melvin
background \rf{mccc} are  equivalent,
being related by a  shift of the R-R field strength parameter:
$b_0= b - R\inv_{11} $.
More generally,  the 
  discussion in \ci{our}
 implies     (assuming again the validity of the 9-11 flip)
            the following generalization of this
relation to the case of the 2-parameter magnetic Melvin model 
 ($R=R_{11}$):
$
{\rm Type\ IIA\ in}\  (\q,\b )\ {\rm   Melvin}$  =
${\rm Type\ 0A\ in}\  (\q -R^{-1},\b )\  {\rm   Melvin}$
 $={\rm Type\ 0A\ in}\  (\q ,\b -\tilde R\inv ) \  {\rm  Melvin  } .$

 It was  conjectured  in \ci{BG}
  that the type 0A tachyon originates from a
massive mode  (in twisted sector of  $\Sigma$-orbifold)
 in  microscopic M-theory,
i.e. a mode which is
absent in the $d=11$  supergravity spectrum.
The  magnetic instability (related to type 0 tachyon)
appears  in different ways in the two models.
In  the first case the tachyon is associated with 
a winding string mode  (or 
winding membrane state in M-theory, see \ci{our}).
In the  second  M-theory
description of type
0A theory -- as M-theory in  the background \rf{surre}
at the critical magnetic field $\b \tilde R_9=1$ 
 the type 0A tachyon  corresponds  to an 
 unstable mode in the   momentum  part of the spectrum  
which may be thus seen directly in $d=10$ or $d=11$ supergravity
spectra  in the corresponding curved backgrounds 
\rf{beb} and \rf{surre}.\footnote{While K-K Melvin background 
is stable as a supergravity solution (the string instability
is associated with a winding mode), 
the  curved T-dual  background \rf{beb} is unstable  already 
at the supergravity level.}
This setting thus  allows one to interpret 
 the type 0A tachyon as 
 a particular fluctuation mode of $d=11$  supergravity
 expanded near \rf{surre}.
 The solution of the relevant  Laplace operatore has the  
 following structure \ci{our}
\be \la{pop}
 M^2=p_9^2+p_{11} ^2 +\b ^2 J^2 + 2
  \b  \sqrt{p_9^2+p_{11}^2}(l_L+l_R+1-S_R+S_L)\ ,
  \ee
  where 
$
J=l_L-l_R+S_L+S_R$  and $p_9= m/R_9, 
\ p_{11} = n/R_{11}$ ($R_9 > R_{11}$).
The tachyon appears for $ l_L=l_R=0, \ S_R = - S_L =1$, 
$m= \pm 1 , \ n=0$.  The corresponding tachyonic mode 
in the  ``dual'' description of type 0 
theory as M-theory on  the flat background \rf{meee}
may be interpreted as a particular wound membrane state \ci{our}
\be 
M^2= (4\pi^2 wR_9R_{11}T_2 ) ^2 +\q ^2 J^2 + 8\pi^2 \q  w  R_9
R_{11}T_2(l_L+l_R+1-S_R+S_L)\  , \ee
where $T_2= (4\pi ^2  R_{11} \a')^{-1} $.
The first term represents the mass  of  a
 membrane of tension $T_2$ wrapped on a
2-torus (wound $w$ times around $x_9$ and
once around $x_{11}$).
The spectrum is similar to \rf{pop},
as expected from the T-duality relation between the corresponding
ten-dimensional models.
The counterpart of the first two terms in \rf{pop} is now a winding
term; the gyromagnetic interaction in \rf{pop}  is traded
for  an analogous  gyromagnetic
term
where the charge is now
the winding number $w$.In this way we determine
 part of the spectrum for the $d=11$ flat model \rf{meee}
 which is relevant for
 the study of instabilities.
The  winding membrane state  that becomes tachyonic at sufficently large $b$ 
has  quantum numbers
$S_R=1=-S_L$, $l_L=l_R=0$.

 \bigskip \bigskip

{\bf Acknowledgments}

\bigskip 

\noindent
I would like to thank the organizers for the excellent
conference  and hospitality.  
The work described in this contribution was done in 
collaboration with J. Russo \ci{our} and 
 was  partially supported by the DOE grant
  DE-FG02-91ER40690 and also by  the  INTAS  99-1590
 and  CRDF RPI-2108 grants.



\bigskip \bigskip
 


\begin{thebibliography}{99}

\bibitem{gib}
G.W.  Gibbons,
``Quantized Flux Tubes In Einstein-Maxwell Theory And
Noncompact Internal Spaces,''
in: {\it Fields and Geometry},
 Proceedings of the 22nd
Karpacz
Winter School of Theoretical Physics, ed.
 A. Jadczyk (World Scientific,
Singapore,  1986).


\bibitem{gaun}
F.~Dowker, J.~P.~Gauntlett, D.~A.~Kastor and J.~Traschen,
``Pair creation of dilaton black holes,''
Phys.\ Rev.\ D {\bf 49}, 2909 (1994)
[hep-th/9309075].
F.~Dowker, J.~P.~Gauntlett, S.~B.~Giddings and G.~T.~Horowitz,
``On pair creation of extremal black holes and Kaluza-Klein
monopoles,''
Phys.\ Rev.\ D {\bf 50}, 2662 (1994)
[hep-th/9312172].
F.~Dowker, J.~P.~Gauntlett, G.~W.~Gibbons and
G.~T.~Horowitz,
``The Decay of magnetic fields in Kaluza-Klein theory,''
Phys.\ Rev.\ D {\bf 52}, 6929 (1995)
[hep-th/9507143].



\bibitem{rohm}
R. Rohm, 
``Spontaneous Supersymmetry Breaking In Supersymmetric String
Theories,''
Nucl.\ Phys.\ B {\bf 237}, 553 (1984).



\bibitem{magnetic}
J.~G.~Russo and A.~A.~Tseytlin,
``Magnetic flux tube models in superstring theory,''
Nucl.\ Phys.\ B {\bf 461}, 131 (1996)
[hep-th/9508068].

\bibitem{closed}
A.~A.~Tseytlin,
``Closed superstrings in magnetic field:
 instabilities and supersymmetry breaking",
Nucl.\ Phys.\ Proc.\ Suppl.\  {\bf 49}, 338 (1996)
[hep-th/9510041].


\bibitem{more}
J.~G.~Russo and A.~A.~Tseytlin,
``Exactly solvable string models of curved space-time backgrounds,''
Nucl.\ Phys.\ B {\bf 449}, 91 (1995)
[hep-th/9502038].
A.~A.~Tseytlin,
``Exact solutions of closed string theory,''
Class.\ Quant.\ Grav.\  {\bf 12}, 2365 (1995)
[hep-th/9505052].





\bibitem{green}
J.~G.~Russo and A.~A.~Tseytlin,
``Green-Schwarz superstring action in a curved magnetic Ramond-Ramond
background,''
JHEP{\bf 9804}, 014 (1998)
[hep-th/9804076].



\bibitem{our} 
J.~G.~Russo and A.~A.~Tseytlin,
``Magnetic backgrounds and tachyonic instabilities in closed 
superstring  theory and M-theory,''
hep-th/0104238.



\bibitem{atick}
J.~J.~Atick and E.~Witten,
``The Hagedorn Transition And The Number
Of Degrees Of Freedom
Of String Theory,''
Nucl.\ Phys.\ B {\bf 310}, 291 (1988).






\bibitem{typ}
L.~J.~Dixon and J.~A.~Harvey,
``String Theories In Ten-Dimensions Without Space-Time
Supersymmetry,''
Nucl.\ Phys.\ B {\bf 274}, 93 (1986).
N.~Seiberg and E.~Witten,
``Spin Structures In String Theory,''
Nucl.\ Phys.\ B {\bf 276}, 272 (1986).


\bibitem{BD}
J.~D.~Blum and K.~R.~Dienes,
``Strong/weak coupling duality relations for non-supersymmetric
string  theories,''
Nucl.\ Phys.\ B {\bf 516}, 83 (1998)
[hep-th/9707160].
 M.~Dine, P.~Huet and N.~Seiberg,
``Large And Small Radius In String Theory,''
Nucl.\ Phys.\ B {\bf 322}, 301 (1989).
  


\bibitem{BG}
O.~Bergman and M.~R.~Gaberdiel,
``Dualities of type 0 strings,''
JHEP{\bf 9907}, 022 (1999) [hep-th/9906055].




\bibitem{CG}
M.~S.~Costa and M.~Gutperle,
``The Kaluza-Klein Melvin solution in M-theory,''
JHEP {\bf 0103}, 027 (2001)
[hep-th/0012072].

\bibitem{GS}
M.~Gutperle and A.~Strominger,
``Fluxbranes in string theory,''
JHEP {\bf 0106}, 035 (2001)
[hep-th/0104136].




\bibitem{sas} 
A.M.~Polyakov,
``String theory as a universal language,''
hep-th/0006132.
``The wall of the cave,''
Int.\ J.\ Mod.\ Phys.\  {\bf A14} (1999) 645
[hep-th/9809057].



\bibitem{KT} 
I.~R.~Klebanov and A.~A.~Tseytlin,
``D-branes and dual gauge theories in type 0 strings,''
Nucl.\ Phys.\ B {\bf 546}, 155 (1999)
[hep-th/9811035].
``A non-supersymmetric large N CFT from type 0 string theory,''
JHEP {\bf 9903}, 015 (1999)
[hep-th/9901101].

\bibitem{TZ} 
A.~A.~Tseytlin and K.~Zarembo,
``Effective potential in non-supersymmetric SU(N) x SU(N) gauge theory  and interactions of type 0 D3-branes,''
Phys.\ Lett.\ B {\bf 457}, 77 (1999)
[hep-th/9902095].

\bibitem{IG} 
I.~R.~Klebanov,
``Tachyon stabilization in the AdS/CFT correspondence,''
Phys.\ Lett.\ B {\bf 466}, 166 (1999)
[hep-th/9906220].

\bibitem{cost}
M.~S.~Costa, C.~A.~Herdeiro and L.~Cornalba,
``Flux-branes and the dielectric effect in string theory,''
hep-th/0105023.


\bibitem{adams} 
A.~Adams and E.~Silverstein,
``Closed string tachyons, AdS/CFT, and large N QCD,''
hep-th/0103220.

\bibitem{ada} 
A.~Adams, J.~Polchinski and E.~Silverstein,
``Don't panic! Closed string tachyons in ALE space-times,''
hep-th/0108075.





\bibitem{KR} 
C.~Kounnas and B.~Rostand,
``Coordinate Dependent Compactifications And Discrete Symmetries,''
Nucl.\ Phys.\ B {\bf 341}, 641 (1990).

\bibitem{KO} 
I.~Antoniadis and C.~Kounnas,
``Superstring phase transition at high temperature,''
Phys.\ Lett.\ B {\bf 261}, 369 (1991).
I.~Antoniadis, J.~P.~Derendinger and C.~Kounnas,
``Non-perturbative temperature instabilities in N = 4 strings,''
Nucl.\ Phys.\ B {\bf 551}, 41 (1999)
[hep-th/9902032].






\end{thebibliography}
\end{document}